\documentclass[twocolumn,showpacs,preprintnumbers,amsmath,amssymb]{revtex4}
\usepackage{graphicx}% Include figure files
\usepackage{dcolumn}% Align table columns on decimal point
\usepackage{bm}% bold math

\begin{document}

\title{
Additional constrains in adsorption-desorption kinetics }% Force line breaks with \\

\author{S.J. Manzi}
\email{smanzi@unsl.edu.ar}
\author{R.E. Belardinelli, G. Costanza}%
\author{V.D. Pereyra}%
\affiliation{%
Departamento de F\'{\i}sica, Instituto de F\'{\i}sica Aplicada (INFAP) - CONICET,
Universidad Nacional de San Luis, Chacabuco 917, 5700
San Luis, Argentina
\\
}%

\date{\today}% It is always \today, today,

\begin{abstract}
In this work, the adsorption-desorption kinetics in the framework of the lattice gas model are analyzed. The transition probabilities are written as an expansion of the occupation configurations. Due to that the principle of detail balance determine half of the adsorption, $A_i$  and desorption, $D_i$ coefficients, different functional relations between them are proposed. Introducing additional constrains, it is demonstrated that when those coefficients are linearly related through a parameter $\gamma$, there are values of lateral interaction, $V$, that lead to anomalous behavior in the adsorption isotherms, sticking coefficients and thermal desorption spectra. Diagrams for the allowed values of $V$ and $\gamma$ are also shown. Alternatively, a non-linear relation among those coefficients is introduced. In such way the equilibrium and non equilibrium observables do not present any anomalous or inconsistent behavior.
\end{abstract}


\maketitle

\section{\label{sec:level1}INTRODUCTION }

The importance of surface and interface in a wide variety of fields has inspired an enormous interest in their structure and dynamics.
The study of the kinetics and dynamics of the surface processes is of fundamental interest in the understanding of heterogeneous catalysis and other processes taking place in gas-solid interfaces [\onlinecite{Cassuto81,Cordoba93,Evans81,Menzel75,Tovbin77,Zhdanov93,Kreuzer86,Kreuzer88a,Kreuzer88b,Payne88,Uebing98,Zhdanov91,Silverberg89,Lombardo91}].
The kinetics is fundamentally determined by the energy transfer \cite{Kreuzer97}.

The rate of change in the adsorbate can be written as a difference between adsorption and desorption terms.
This can be made in a heuristic way or in the framework of a rigorous approach using the non-equilibrium thermodynamic theory [\onlinecite{Kreuzer86,Kreuzer96,Kreuzer97}].

If the adsorbate does not remain in a quasi-equilibrium state during desorption, then the description of the system through macroscopic variables is not enough and it should be carried out based on the non-equilibrium statistical mechanics that involves time-dependent distribution functions. The Kinetic Lattice Gas Model (KLGM) is an example of such theory, and it was set up in close analogy to the time dependent Ising model for magnetic systems, which was originally introduced by Glauber [\onlinecite{Glauber63,Kawasaki72}]. In its simplest form the KLGM is restricted to the submonolayer regime and to the gas-solid system where the surface structure and the adsorption sites do not change with the coverage \cite{Kreuzer90}. However, further generalizations of the lattice-gas model have been made in the past, for example: the multilayer lattice gas model was used for analyzing and simulating desorption data for Li and Cs on Ru(0001) [\onlinecite{Payne96,Payne02a}]; the influence of the precursor state in the adsorption-desorption kinetics \cite{Kreuzer96} and other related problems \cite{Payne02b}; the adsorption-desorption kinetics with multiple site occupations [\onlinecite{Costanza03,Furlong04,Costanz06}], etc.

On the other hand, it is well known that different microscopic dynamics can yield different equilibrium paths and equilibrium fluctuations \cite{Schochet92} (cluster versus local MC algorithms being the most extreme example \cite{Landau}) and even noticeable differences in the steady-state microstructure [\onlinecite{Siegert94,Shim01}]. Nevertheless, the general expectation is that, if no additional parameters (such as an activation barrier or a diffusion rate) are introduced into the physical model, observables are only affected quantitatively.

Recent studies indicate that different stochastic dynamics, even when they have the same conserved quantities and
satisfy detailed balance, lead to important differences in the nanostructure of field-driven
interfaces [\onlinecite{Rikvold00,Rikvold02a,Rikvold02b,Rikvold03,Buendia06}].

Before defining the transition probabilities, an important distinction must be made. That is, between models with hard dynamics[\onlinecite{Marro99,Rikvold02b,Buendia06,Frank05,Saito00,Park02,Buendia07}], in which the single-site transition rates cannot be factorized into one term that depends only on the interaction energies, in contrast with those models with soft dynamics [\onlinecite{Buendia06,Kang89,Fichthorn91,Rikvold02a,Rikvold03,Frank06,Buendia04,Park04,Ala92a,Ala92b}], for which this factorization is possible.

In this context, Kang et al. \cite{Kang89} have analyzed different choices for the transition probabilities in MC simulations for studying the growth exponent in the domains growth. They showed that the choice of transition probabilities affects directly the dynamic quantities. Particularly, they found this dependence in the growth exponent in the ordered domains out of equilibrium \cite{Kawasaki66}.
In the same way, Rikvold et al. have shown that the intrinsic interface width and properties in field-driven solid-on-solid interface studies depend on the choice of dynamics. They have found that, in the framework of the soft dynamics, all dependence on the field canceled due to the principle of detailed balance [\onlinecite{Rikvold02a,Rikvold03,Buendia06}]. On the other hand in hard dynamics the intrinsic interface width and properties, such as the propagation velocity, are strongly affected by the field \cite{Rikvold02a}.

Despite the fact that, detailed microscopic mechanisms of the surface processes are usually not known, in the KLGM the transition probabilities can be written in terms of the occupation configurations of all neighboring sites [\onlinecite{Wierzbicki91,Kreuzer86,Kreuzer96,Kreuzer97,Kreuzer90,Payne96,Payne02a,Payne02b,Payne07}]. In this point, the principle of detailed balance imposes a set of restrictions on the coefficients of adsorption $A_i$, desorption $D_i$, diffusion $C_i$, etc., as is discussed by H. J. Kreuzer and co-workers [\onlinecite{Kreuzer96,Kreuzer97,Kreuzer90,Payne02b}], where the authors have introduced different kinetics according to the relations among those coefficients. When adsorption and desorption are the only processes taken into account, they have assumed linear relation $A_i=\gamma D_i$ (where $\gamma$ is a proportionality parameter). In such way the authors have obtained the adsorption-desorption kinetics calculating the sticking coefficients for different cases, including Langmuir kinetics, sticking on random adsorbate and the influence of intrinsic and extrinsic precursors \cite{Kreuzer96}. Moreover, very recently, S. H. Payne and H. J. Kreuzer have discussed the one-dimensional diffusion with different lateral interaction, using the same methodology \cite{Payne07}.

Although those coefficients satisfy the principle of detailed balance, the relation between them depends strongly on the lateral interaction energy and must be carefully analyzed according to the adsorption and desorption processes, otherwise the results could be not general leading to anomalous behaviors in the kinetics.

The aim of the present paper is to demonstrate that, even for the simplest one-dimensional case, a linear relation among the adsorption and desorption coefficients gives some inconsistencies. With this purpose, the adsorption desorption kinetics are analyzed for a 1D KLGM with nearest neighbor lateral interaction, $V$.

The analytical treatment have been done in a one-dimensional system, mainly due to two reasons. The first one, is the availability of the exact solution for the coverage and two-sites correlation function (which is not possible in higher dimensions). The second one, is the presence of phase transitions in two and higher dimensions. This could mask the possible anomalous behaviors in the observables, which is precisely the objective of the present investigation.

The outline of the paper is as follows: In Section II, the KLGM is set up, introducing the master equation and writing the transition probabilities in terms of the occupation configurations of all neighboring sites. A set of restrictions on the adsorption and desorption coefficients are obtained according to the principle of detailed balance. In order to fix the unknown coefficients in the transition probabilities, two different linear relations are proposed. The equilibrium and non equilibrium properties of the system are analyzed solving the rate equations for the coverage and higher correlations. Additional inequalities among the adsorption and desorption parameters are found, determining forbidden values for the parameters $\gamma$ and $V$, for which the adsorption isotherms, sticking coefficients and thermal desorption spectra are ill-behaved. Diagrams with forbidden regions as function of lateral interactions $V$ and the parameter $\gamma$ are showed. In order to solve these inconsistencies, in Section IV, a functional relation between the adsorption and desorption parameters is introduced. In such way the observables present a correct behavior without inconsistencies. Finally, in Section V our conclusions are presented.

\section{\label{sec:level1}THE KINETIC LATTICE GAS MODEL AND THE MASTER EQUATION }

To set up the KLGM one restricts the analysis to a gas-solid system in which all relevant processes, like diffusion, adsorption, desorption, reactions, etc., are Markovian. One assumes that the system can be divided into cells, labeled $i$, for which one introduces microscopic variables $n_i=1$ or $0$ depending on whether cell $i$ is occupied by an adsorbed gas particles or not. The connection with magnetic systems is made by a transformation to spin variables $\sigma_i=2n_i-1$. To introduce the dynamic of the system one writes down a model Hamiltonian

\begin{eqnarray}
H=E_s\sum_i n_i +\frac{1}{2}V\sum_{\langle ij \rangle} n_in_j+...
\end{eqnarray}

Here $E_s$ is a single particle energy, $V$ is the two particle interaction between nearest neighbors $\langle ij \rangle$. Interactions between next nearest neighbors etc., and many particle interactions can be easily added to eq. (1).

As long as the number of particles in the adsorbate does not changed, which is the case for systems in equilibrium or diffusion studies, the first term in (1) is constant and can be dropped from further consideration. However, if the objective is the study adsorption-desorption kinetics, the number of particles in the adsorbate changes as a function of time and a proper identification of $E_s$ is mandatory. Arguing that the lattice gas Hamiltonian should give the same Helmholtz free energy as a microscopic Hamiltonian (for noninteracting particles) one can show that the proper identification is given by \cite{Kreuzer99},

\begin{eqnarray}
E_s=-V_0-\frac{1}{\beta} \ln(q_3 q_{int}) -\frac{1}{\beta}\left[\ln(\beta\lambda^3P)-\ln(Z_{int})\right]
\end{eqnarray}

where $\beta = 1/k_BT$; $k_B$ and $T$ are the Boltzmann constant and the absolute temperature, respectively. $V_0$ is the (positive) depth of the surface potential, $q_3$ is the single particle partition function of an adsorbed particle, $q_{int}$ is the internal partition function for (frustrated) vibrations and rotations of the adsorbed molecule; $P$ is the pressure in the gas phase above the surface and $\lambda=h/\sqrt{2\pi m k_BT}$ is the thermal wavelength of the adparticle with mass $m$, which partition function for (free) vibrations and rotations is $Z_{int}$.

One introduces a function $P({\bf n},t)$ which gives the probability that a given microscopic configuration ${\bf n}= (n_1,n_2,...,n_N)$ is realized at time $t$, where $N$ is the total number of adsorption sites on the surface. It satisfies a master equation,

\begin{eqnarray}
dP({\bf n};t)=\sum_{n\acute{}}\left[W({\bf n};{\bf n\acute{}})P({\bf n\acute{}};t)-W({\bf n\acute{}};{\bf n})P({\bf n};t)\right]
\end{eqnarray}

where $W({\bf n\acute{}};{\bf n})$ is the transition probability that the microstate $\bf n$ changes into $\bf n\acute{}$ per unit time. It satisfies detailed balance

\begin{eqnarray}
W({\bf n\acute{}};{\bf n})P_0({\bf n})=W({\bf n};{\bf n\acute{}})P_0({\bf n\acute{}})
\end{eqnarray}

where

\begin{eqnarray}
P_0({\bf n})=  Z^{-1} \exp\left(-\beta H({\bf n})\right)
\end{eqnarray}

is the equilibrium probability and $Z$ is the partition function given by

\begin{eqnarray}
Z=\sum_{\lbrace\bf n \rbrace} e^{-\beta H(\bf n)}
\end{eqnarray}

In principle, $W({\bf n\acute{}};{\bf n})$ must be calculated from a Hamiltonian that includes, in addition to (1), coupling terms to the gas phase and the solid that mediate mass and energy exchange. However, depending of the system, different expressions for transition probabilities can be proposed. In transition dynamic approximation [\onlinecite{Ala92a,Buendia04}], transition rates cannot be factorized into one part that depends only on the interaction energy and another that depends only on the field energy \cite{Buendia06} (hard dynamic \cite{Marro99}).

Usually, one follows the procedure introduced by Glauber and guesses an appropriate form for $W({\bf n\acute{}};{\bf n})$. One further assumes that the duration of an individual transition, e.g., hopping to a neighboring site, is much shorter than the residence time in the initial state. In this situation there will be only one transition at any given time and  the total transition probability as a sum of individual terms can be written.

In order to analyze the simplest cases, let us consider the one-dimensional lattice gas with nearest-neighbors interactions where only direct adsorption and desorption processes are taken into account (no other processes are considered), in this case, the transition probability can be written as,

\begin{eqnarray}
&&W_{ad-des}({\bf n\acute{}};{\bf n})=\sum_{i}[w_a(1-n_i)(A_0+A_1(n_{i-1}+n_{i+1})\nonumber\\
&&+A_2n_{i-1}n_{i+1})+w_dn_i(D_0+D_1(n_{i-1}+n_{i+1}) \nonumber\\
&&+D_2n_{i-1}n_{i+1})] \delta(n\acute{}_i,1-n_i)\Pi_{j\neq i}\delta(n\acute{}_j,n_j)
\end{eqnarray}

Here adsorption into site $i$ occurs if $n_i=0$ initially, with a rate controlled by prospective neighbors if $A_i\neq 0$. The Kronecker delta for sites $j\neq i$ excludes multiple transitions.

The motion equation for coverage can be obtained by multiplying the master equation by occupation number $n_i$ and summing overall sites. Thus one can obtain the following expression for the coverage [\onlinecite{Kreuzer88a,Kreuzer88b,Payne88,Wierzbicki91,Costanza03,Payne93}]

\begin{eqnarray}
\frac{d\theta}{dt}&=& w_a \left[A_0 E+2 A_1 NE+ A_2 NEN\right]\nonumber\\
&&- w_d \left[D_0 N+2 D_1 NN+ D_2 NNN\right]
\end{eqnarray}

With the same procedure the rate equations for the next correlation functions for three independent sites can be written as,

\begin{widetext}
\begin{eqnarray}
\frac{dNN}{dt}&=&2w_a \left[(A_0+A_1) NE+ (A_1+A_2) NEN\right]- 2w_d\left[(D_0+D_1) NN+ (D_1+D_2) NNN\right]\\
\frac{dNEN}{dt}&=&w_a \left[2A_0 NEE + 2A_1 NEEN - (A_0+2A_1+A_2) NEN\right]\nonumber\\
&&- w_d \left[2D_0 NEN + 2D_1 NNEN + (D_0+2D_1+D_2) NNN\right]\\
\frac{dNNN}{dt}&=& w_a \left[2(A_0+A_1) NNE + (A_0+2A_1+A_2) NEN + 2(A_1+A_2) NENN\right]\nonumber\\
&&- w_d\left[(3D_0+4D_1+D_2) NNN+ 2(D_1+D_2) NNNN\right].
\end{eqnarray}
\end{widetext}

The principle of detailed balance imposes a set of restrictions on the coefficients $A_i$ and $D_i$ \cite{Payne02b}, which are:

\begin{eqnarray}
w_aA_0=w_dD_0e^{-\beta E_s}
\end{eqnarray}

\begin{eqnarray}
w_a(A_0+A_1)=w_d(D_0+D_1)e^{-\beta (E_s+V)}
\end{eqnarray}

and

\begin{eqnarray}
&&w_a(A_0+2A_1+A_2)=w_d(D_0+2D_1+D_2)e^{-\beta (E_s+2V)}\nonumber\\
&&
\end{eqnarray}

$w_a$ and $w_d$ cannot be fixed by detailed balance, because they contain the information about the energy exchange with the solid in the adsorption and desorption processes, which is not in the static lattice gas Hamiltonian \cite{Payne02b}. However, if one consider that $w_a=w_d=w_0$, and by comparison with the phenomenological expression for adsorption, one can identify it as

\begin{eqnarray}
w_0=S_0(T)\frac{P\lambda a_s}{h},
\end{eqnarray}

$S_0(T)$ is the temperature-dependent sticking coefficient at zero coverage and $a_s$ is the area of a surface unit cell. It contains the dynamic information about the energy transfer from the adsorbing particle to the solid which gives rise to its temperature dependence, for instance, an exponential Boltzmann factor for activated adsorption. It can be calculated only on the basis of a dynamic theory that accounts for the coupling of the ad-particles to the vibrational and electronic degrees of freedom of the substrate, and must be postulated ad hoc within the context of the kinetic lattice gas model.

The long-range interaction introduces similar constraints on the other coefficients. Each such constraints introduces two new coefficients. However, detailed balance provides only half the number of relations to fix these unknown coefficients in the transition probabilities. Again, the static (lattice gas) Hamiltonian cannot completely dictate the kind of kinetics possible in the system. As it is pointed out in references [\onlinecite{Kreuzer97,Kreuzer90,Payne02b}], any functional relation between the $A-$ and $D-$ coefficients must be postulated ad hoc, or calculated from a microscopic Hamiltonian that accounts for coupling of the adsorbate to the lattice or electronic degrees of freedom of the substrate.

In the next section two different relations among the adsorption and desorption coefficients, are analyzed.

\section{\label{sec:level1}LINEAR RELATIONS BETWEEN ADSORPTION AND DESORPTION COEFFICIENTS }

A linear relationship between the $A_i$ and $D_i$ coefficients is proposed (case K1) as:

\begin{eqnarray}
\frac{A_i}{A_0}=\gamma \frac{D_i}{D_0},
\end{eqnarray}

for $i=1,2$, where $\gamma$ is a proportionality coefficient. Note that this linear relation was previously proposed in references [\onlinecite{Kreuzer97,Kreuzer90,Payne02b}], where $A_0=1$ has been considered.
The principle of detailed balance imposes,

\begin{eqnarray}
D_1=D_0 \left( \frac{e^{\beta V}-1}{1-\gamma e^{\beta V}} \right)
\end{eqnarray}

and

\begin{eqnarray}
D_2=D_0 \left( \frac{e^{2\beta V}-1}{1-\gamma e^{2\beta V}}\right)-2D_1
\end{eqnarray}

Other possible choice of the parameters $A_i$ and $D_i$, which fulfill the detailed balance and leads to a new formulation of the adsorption-desorption kinetics, is (case K2),

\begin{eqnarray}
A_{2}=\gamma A_{1}
\end{eqnarray}

and

\begin{eqnarray}
D_{2}=\gamma D_{1}
\end{eqnarray}

Considering only nearest neighbor interactions one can replace (19) and (20) in the equations (13) and (14) and find the following expressions for the coefficients:

\begin{eqnarray}
A_1=A_0\left( \frac{(2+\gamma)e^{-\beta V}-(1+\gamma)e^{-2\beta V}-1}{(2+\gamma)(1-e^{-\beta V})}\right)
\end{eqnarray}

and

\begin{eqnarray}
D_1=D_0\left( \frac{(1+\gamma)e^{\beta V}-(2+\gamma)+e^{-\beta V}}{(2+\gamma)(1-e^{-\beta V})}\right).
\end{eqnarray}

When adsorption and desorption coefficients are determined, the properties of the system such as adsorption isotherms, sticking coefficient and thermal desorption spectra can be obtained.

However, to analyze correctly the consequences of the linear relations assumed above, let us introduce the following identities \cite{ben-Avraham92}

\begin{eqnarray}
E=NEN+2NEE+EEE
\end{eqnarray}

and

\begin{eqnarray}
NE=NEE+NEN
\end{eqnarray}

In such way, one can easily obtain alternative expressions for the equations (8)-(11) as,

\begin{widetext}
\begin{eqnarray}
\frac{d\theta}{dt}&=& w_0\left[(A_0) EEE+2(A_0+A_1)NEE+(A_0+2A_1+A_2)NEN\right]\nonumber\\
&&-w_0\left[(D_0) ENE+2(D_0+D_1)NNE+(D_0+2D_1+D_2)NNN\right]\\
\frac{dNN}{dt}&=& 2w_0\left[(A_0+A_1)NEE+(A_0+2A_1+A_2)NEN\right]\nonumber\\
&&-2w_0\left[(D_0+D_1)NNE+(D_0+2D_1+D_2)NNN\right]\\
\frac{dNEN}{dt}&=& w_0\left[2(A_0) NEEE+2(A_0+A_1)NEEN-(A_0+2A_1+A_2)NENN-(A_0+2A_1+A_2)NENE\right]\nonumber\\
&&-w_0\left[2(D_0) NENE+2(D_0+D_1)NENN-(D_0+2D_1+D_2)NNNE-(D_0+2D_1+D_2)NNNN\right]\nonumber\\
\\
\frac{dNNN}{dt}&=&w_0\left[2(A_0+A_1)NNEE+3(A_0+2A_1+A_2)NENN+(A_0+2A_1+A_2)NENE\right]\nonumber\\
&&-w_0\left[(2(D_0+D_1)+(D_0+2D_1+D_2))NNNE+3(D_0+2D_1+D_2)NNNN\right]
\end{eqnarray}
\end{widetext}

Note that $EEE$, $NEE$, $NEN$, $ENE$, $NNE$ and $NNN$, as well as $NEEE$, $NEEN$, $NENN$, $NENE$, $NNNE$, $NNNN$ and $NNEE$ are mutually exclusive conditional probabilities, therefore each of the parentheses in eqs. (25)-(28) must be positive. After some algebra one can obtain the following inequalities

\begin{eqnarray}
Q^0_{ads}=A_0\geq 0,
\end{eqnarray}

\begin{eqnarray}
Q^1_{ads}=(A_0+A_1)\geq 0,
\end{eqnarray}

\begin{eqnarray}
Q^2_{ads}=(A_0+2A_1+A_2)\geq 0,
\end{eqnarray}

\begin{eqnarray}
Q^0_{des}=D_0\geq 0,
\end{eqnarray}

\begin{eqnarray}
Q^1_{des}=(D_0+D_1)\geq 0
\end{eqnarray}

and

\begin{eqnarray}
Q^2_{des}=(D_0+2D_1+D_2)\geq 0.
\end{eqnarray}

It it easy to see that eqs. (29)-(34) impose new restrictions on the $A_i$ and $D_i$ coefficients.
If these additional restrictions are applied to the two proposed kinetics, certain values of parameter $\gamma$ are not allowed. This forbidden values depend on the lateral interaction $V$, and this is shown in the diagrams of the Figs. 1a) y 1b) for the K1 y K2 kinetics, respectively.

Region I corresponds to those values of $\gamma$ and $V$ which make positive all the coefficients in eqs. (29)-(34). Region II corresponds to those values of parameters which make $Q^1_{ads}\geq 0$, $Q^1_{des}\geq 0$, $Q^2_{ads}< 0$ and $Q^2_{des}< 0$ (intermediate region). Region III corresponds to those values of $\gamma$ and $V$ which make all the coefficients negative, $Q^1_{ads} < 0$, $Q^2_{ads} < 0$, $Q^1_{des} < 0$ and $Q^2_{des} < 0$ (forbidden region).

\begin{figure}
\includegraphics[scale=0.4]{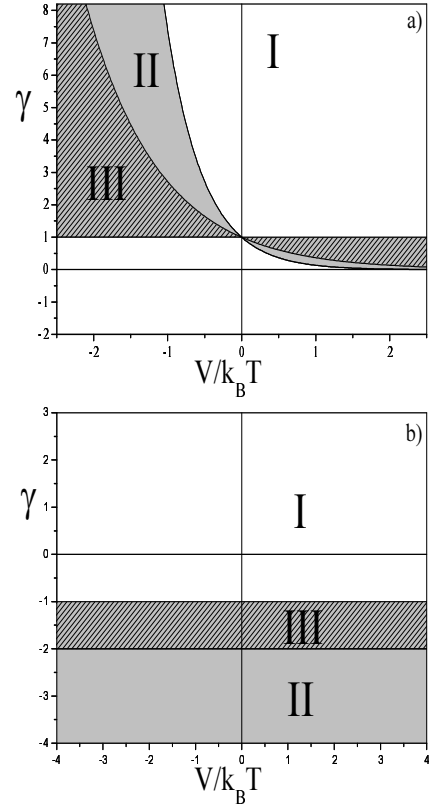}% Here is how to import EPS art
\caption{\label{fig:epsart} Diagram $\gamma$ versus $V/k_BT$ corresponding to a) K1 and b) K2 proposed kinetics.}
\end{figure}

To calculate the exact solution for adsorption isotherm the two first rate equations (eqs. (25) and (26)) must be set equal to zero [\onlinecite{Payne93,Costanza03}].

To calculate the sticking coefficient one used the following definition \cite{Kreuzer96}:

\begin{eqnarray}
S(\theta,T)= A_0 E+2 A_1 NE+ A_2 NEN
\end{eqnarray}

The exact solution for immobile TPD spectra is obtained by solving the rate equations for the first four correlations ((25)-(28)), where the adsorption terms are neglected [\onlinecite{Evans87, Payne93,Costanza03}]. In all TPD spectra  the thermal desorption is considered as an activated process. Where the activation energy is $10$ kcal/mol, the pre-exponential factor is $10^{13} s^{-1}$ and the initial coverage is $\theta_0=0.9$.

The adsorption isotherms, sticking coefficients and TPD spectra for immobile adsorbate are showing in Figs. 2, 3 and 4, respectively.

For those values of the parameter $\gamma$ out of the allowed region, the first derivative of the isotherms presents a discontinuity as a function of the chemical potential, and both the sticking coefficient as TPD spectra take negative values. These behaviors are certainly anomalous.

\begin{figure}
\includegraphics[scale=0.4]{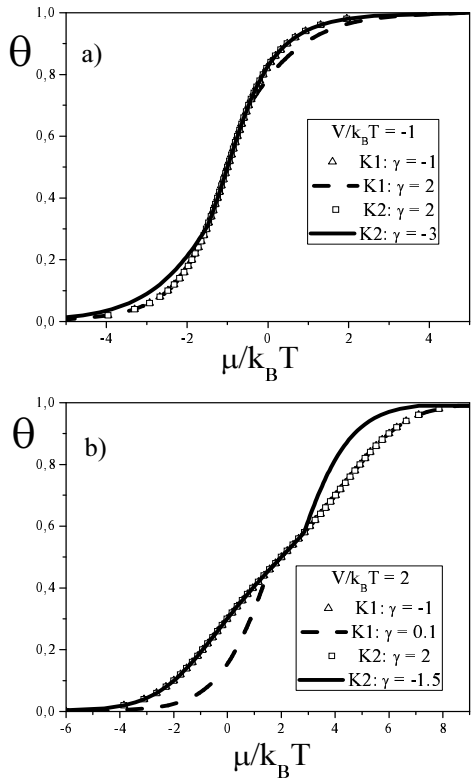}% Here is how to import EPS art
\caption{\label{fig:epsart} Adsorption isotherms for different values of the parameter $\gamma$ and for two linear proposed kinetics. a) Attractive and b) Repulsive lateral interaction.}
\end{figure}

\begin{figure}
\includegraphics[scale=0.4]{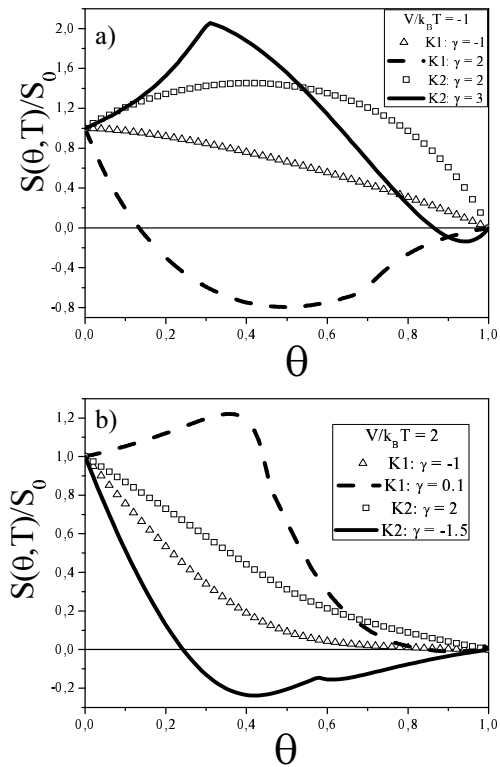}% Here is how to import EPS art
\caption{\label{fig:epsart} Sticking coefficients for different values of the parameter $\gamma$ and for two linear proposed kinetics. a) Attractive and b) Repulsive lateral interaction.}
\end{figure}

\begin{figure}
\includegraphics[scale=0.4]{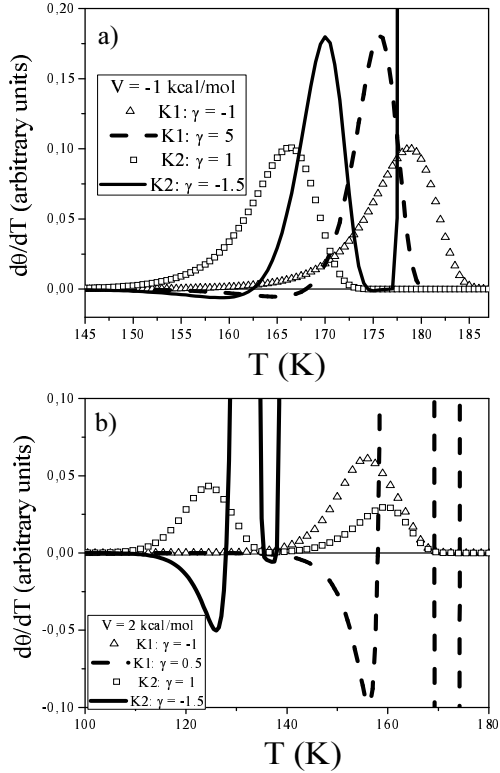}% Here is how to import EPS art
\caption{\label{fig:epsart} Immobile TPD spectra for different values of the parameter $\gamma$ and for two linear proposed kinetics. a) Attractive and b) Repulsive lateral interaction.}
\end{figure}

For values of $\gamma$ belonging to the allowed region the behavior of the observable do not present any inconsistencies. Although the physical meaning of the relations given in eqs. (16), (19) and (20) is not clear, it seem to be perfectly valid in view that they fulfill the detailed balance. However, it must be certain coherence in the behavior of the observables according to the lateral interactions. For example, for repulsive lateral interaction, the adsorption isotherms for monomers present a characteristic plateau at coverage $\theta=1/2$ for enough low temperature. This is due to an ordering of the adsorbed particles, namely, there is an alternation between particles and vacancies. For the same interaction the sticking coefficient will be below the straight line  $S(\theta)=1-\theta$, which corresponds to the Langmuir kinetics (null interaction) \cite{Kreuzer96}. Finally, the immobile TPD spectra for high initial coverages will have three peaks, as is expected [\onlinecite{Sales87,Sales96}]. This behavior can not be reproduced by using both proposals, K1 and K2.

Moreover, in the framework of proposal K1, the three peaks in the immobile TPD spectra can be observed only in the limit $\gamma \rightarrow 0$. However, this value of $\gamma$ corresponds to the Langmuir kinetics.

Clearly, the principle of detailed balance is not enough to guarantee the correct behavior of the kinetics and deeper analysis must be done to choose the functional relation between the $A-$ and $D-$ coefficients.

\section{\label{sec:level1}INVERSE RELATION BETWEEN THE ADSORPTION AND DESORPTION COEFFICIENTS }

As is discussed in the last section, the relation among the adsorption and desorption coefficients must be carefully chosen. In particular, linear relations between them, as is proposed by Kreuzer and co-worker in ref. [\onlinecite{Kreuzer97,Kreuzer90,Payne02b}] are not valid in general and some values of the parameter $\gamma$ and $V$ are not allowed. In fact, the observables obtained for those values of $\gamma$ and $V$ are ill-behaved. In order to solve this kind of inconsistence, let us introduced the following relations between the $A-$ and $D-$ coefficients (K3),

\begin{eqnarray}
A_0=\frac{1}{D_0},
\end{eqnarray}

\begin{eqnarray}
A_0+A_1=\frac{1}{D_0+D_1}
\end{eqnarray}

and

\begin{eqnarray}
A_0+2A_1+A_2=\frac{1}{D_0+2D_1+D_2}.
\end{eqnarray}

Using the detailed balance expressed in eqs. (12)-(14) and after some algebra, one can obtain the following expression for the $A-$ and $D-$ coefficients:

\begin{eqnarray}
A_0=\exp[-\beta E_s/2]
\end{eqnarray}

\begin{eqnarray}
D_0=\exp[\beta E_s/2]
\end{eqnarray}

\begin{eqnarray}
A_1=\exp[-\beta E_s/2](\exp[-\beta V/2]-1)
\end{eqnarray}

\begin{eqnarray}
D_1=\exp[\beta E_s/2](\exp[\beta V/2]-1)
\end{eqnarray}

\begin{eqnarray}
A_2=\exp[-\beta E_s/2](\exp[-\beta V/2]-1)^2
\end{eqnarray}

and

\begin{eqnarray}
D_2=\exp[\beta E_s/2](\exp[\beta V/2]-1)^2.
\end{eqnarray}

Then, one can write the adsorption-desorption coefficients as follow:

\begin{eqnarray}
Q_{ads}^i=\exp[-\beta (E_s/2+iV/2)],
\end{eqnarray}

and

\begin{eqnarray}
Q_{des}^i=\exp[\beta (E_s/2+iV/2)],
\end{eqnarray}

with $i=0,1,2$. These expressions for the coefficients allow an adsorption-desorption kinetics without physical inconsistencies.  Particularly, in the framework of this formulation, Langmuir kinetics arises only for $V=0$. In Figure 5 a), b) and c) it is shown the adsorption isotherms, sticking coefficients and TPD spectra for different lateral interaction using the above formulation. As is observed, there are coherence between the behavior of the observables.

\begin{figure}
\includegraphics[scale=1]{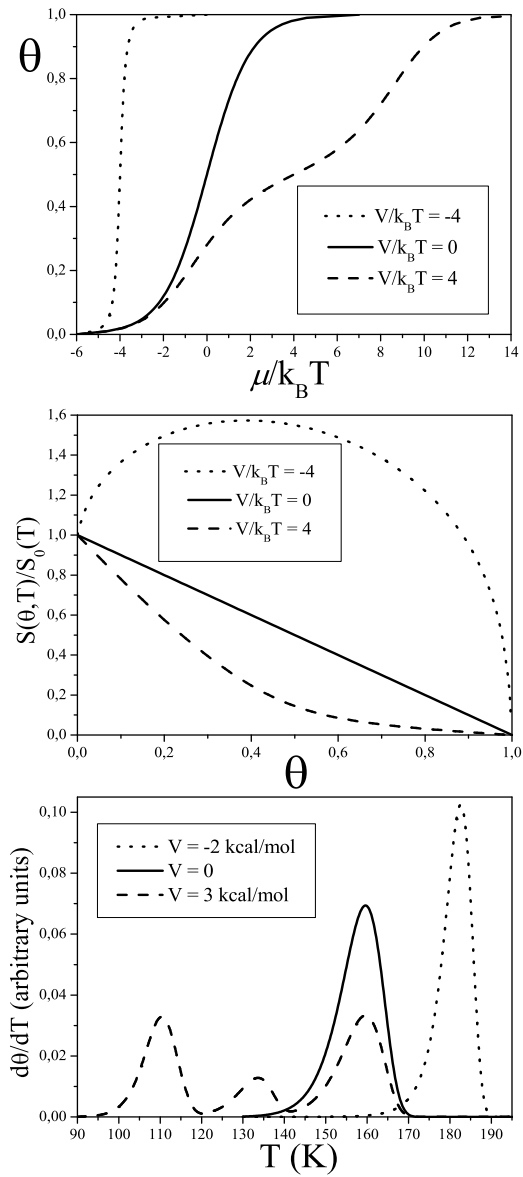}% Here is how to import EPS art
\caption{\label{fig:epsart} a) Adsorption isotherms, b) sticking coefficients and c) immobile TPD spectra, for  attractive (dot), repulsive (dash) and null (line) interaction, using the inverse relation L3}
\end{figure}

\section{\label{sec:level1} SUMMARY AND CONCLUSIONS }

In this work, the adsorption-desorption kinetics in the framework of the one-dimensional lattice gas model with nearest neighbor lateral interaction is considered. The master equation approach has been used to derive the rate equations for coverage and higher correlations, which gives the equilibrium and non-equilibrium properties of the system.
In order to obtain those equations, transition probabilities are written in terms of the occupation configurations. In such way, the principle of details balance imposes a set of restrictions on the adsorption, $A_i$ and desorption, $D_i$ coefficients. However it determines half of them. To overcome such difficulty different functional relations among those coefficients can be postulated, the simplest one is a linear relation. However, it is observed that, when rate equations are written in terms of mutually exclusive conditional probabilities, additional constrains appear on $A_i$ and $D_i$ coefficients. These are rather restrictive because not all the values of the lateral interactions, $V$ and linear parameter, $\gamma$ are allowed.
Moreover, the kinetics obtained for some values of these parameters are wrong, or at least, the equilibrium and non-equilibrium observables present anomalous behavior.
A diagrams of the allowed and forbidden values as a function of the parameters $V$ and $\gamma$, as well as the adsorption isotherms, sticking coefficient and TPD spectra for two different linear relations are showed. Three well-defined regions emerge from these diagrams in both cases: (i) The allowed region, where the adsorption-desorption kinetics is well behaved, i.e., the isotherms, sticking coefficients and the TPD spectra are well behaved. It is also observed that Langmuir ($A_i=0$) and interaction kinetics ($A_i =- D_i$) belong to this region. (ii) The intermediate region, where the first derivative of the adsorption isotherms are discontinuous. This results is not possible in a one-dimensional lattice gas with nearest-neighbor lateral interaction. (iii) The forbidden region, where the adsorption isotherms behave like in region II and sticking coefficients are negative.
In order to solve such inconsistences, an inverse relation among adsorption and desorption coefficients is introduced. In such way the equilibrium and non equilibrium observables do not present any anomalous or inconsistent behavior.

As a general conclusion, the principle of detailed balance is not enough to guarantee the correct behavior of the kinetics. The linear relations among the adsorption and desorption coefficients are not general and some values of the lateral interaction $V$ and $\gamma$ are not allowed. Inverse relations among the adsorption and desorption coefficients seems to be a better choice to determine the rate equations. However, deeper analysis must be done in order to choose a more general functional relation between the $A-$ and $D-$ coefficients.

On the other hand, both linear relations as well the inverse one belong to the so called soft dynamics, where transition probabilities factorize into a part due only to the change in the field energy and a part due only to the change in the interaction energy.

Finally, it is important to emphasize that, even when the treatment has been done in one-dimensional system, the results seems to be general and they do not depend on the dimensionality. However, the extension to higher dimensions, the incorporation of next-nearest neighbor interactions, as well as of diffusion terms should be done in next future.

\begin{acknowledgments}

We thank Prof. G. Zgrablich and Dr. M.R. G\'{o}mez for reading the manuscript. This work is partially supported by the CONICET (Argentina).
\end{acknowledgments}

\newpage %Just because of unusual number of tables stacked at end

\end{document}